\documentclass[aps,pra,showpacs,groupedaddress,twocolumn]{revtex4}
\usepackage{amsfonts}
\usepackage{graphicx}
\usepackage{latexsym}
\usepackage{makeidx}
\usepackage{amsmath}
\usepackage{amssymb}
\usepackage{graphicx}

\begin{document}

\title{Dynamical Casimir Effect in two-atom cavity QED}
\author{A. V. Dodonov and V. V. Dodonov}
\affiliation{Instituto de F\'{\i}sica, Universidade de Bras\'{\i}lia, PO Box 04455,
70910-900, Bras\'{\i}lia, Distrito Federal, Brazil}

\begin{abstract}
We study analytically and numerically the dynamical Casimir effect in a
cavity containing two stationary 2-level atoms that interact with the
resonance field mode via the Tavis--Cummings Hamiltonian. We determine the
modulation frequencies for which the field and atomic excitations are
generated and study the corresponding dynamical behaviors in the absence of
damping. It is shown that the two-atom setup allows for monitoring of photon
generation without interrupting the growth, and different entangled states
can be generated during the process.
\end{abstract}

\maketitle

\emph{Introduction.}--In view of the recent progress \cite{DCE-Nature} in
experiments on the observation of the so called Dynamical Casimir Effect
(DCE) \cite{revRMP}, the problem of detecting photons generated from the
initial vacuum state
becomes quite actual.
It was shown long ago \cite{PLA} that the presence of a detector can
change significantly the statistics (including the mean number) of created
quanta, compared with the predictions made for an idealized empty cavity model.
Therefore it is necessary to study in detail different detection schemes.
At least two main schemes were proposed until now. In the so called
MIR experiment the quanta of the microwave field are to be detected by an
antenna put inside the closed cavity \cite{Padua05}. Another idea was
to use as detectors real Rydberg atoms passing through the cavity
\cite{PLA,shake,Onof06,Kawa11}
or ``artificial atoms'' \cite{arx1} in the case of Circuit QED systems,
such as those described in \cite{DCE-Nature,revRMP}.
The simplest solutions for a detector
modeled as a single two-level atom were obtained in \cite{PLA,shake}, and
recently results of more detailed theoretical and numerical studies of the
atom-field interaction during the DCE were presented in \cite%
{Roberto,Resdyn,Resact}. Three-level models of detectors were considered in
\cite{Pero11,Tridin}.
It seems important to analyze different configurations
to choose the optimal scheme.

Here we study how
the DCE dynamics is affected by the presence of
\emph{two\/} 2-level atoms (detectors) interacting with a single resonance
cavity field mode. Our starting point is the Hamiltonian (we set $\hbar =1$)%
\[
H_{0} =\omega _{t}n+\sum_{j=1}^{2}\Big[ \frac{\Omega _{j}}{2}\sigma
_{j}^{z}+g_{j}(a\sigma_{j}^{+}+a^{\dagger }\sigma_{j}^{-})\Big]
-i\chi _{t}( a^{2}-a^{\dagger 2})
\]
where $a$ ($a^{\dagger }$) is the cavity annihilation (creation) operator
and $n\equiv a^{\dagger }a$ is the photon number operator.
The Pauli operators are defined as $\sigma _{j}^{z}=|%
\overline{e}_{j}\rangle \langle \overline{e}_{j}|-|\overline{g}_{j}\rangle
\langle \overline{g}_{j}|$, $\sigma _{j}^{-}=|\overline{g}_{j}\rangle
\langle \overline{e}_{j}|$, $\sigma _{j}^{+}=|\overline{e}_{j}\rangle
\langle \overline{g}_{j}|$, where $|\overline{g}_{j}\rangle $ and $|%
\overline{e}_{j}\rangle $ are the ground and excited states of
the $j$-th atom ($j=1,2$), respectively. $\Omega _{j}$ and $g_{j}$ are the
atomic transition frequencies and the atom-field coupling constants (assumed
real for simplicity).
If $\chi _{t}=0$, then $H_0$ is the special case of the known
Tavis--Cummings Hamiltonian \cite{Tavis} studied in numerous papers (see,
e.g., \cite{Saav98,Vad03,Tess03,Garr} and references therein). Physical
realizations of this Hamiltonian
(which holds for $|g_j| \ll \Omega_j$)
were demonstrated in \cite{Retz07} for
trapped ions and in \cite{Fink09} for the Circuit QED systems.

The last term in $H_0$ describes the effect of photon creation
(equivalent to squeezing) in a cavity whose fundamental eigenfrequency
varies in time due to the motion of a boundary \cite{PLA,Law94,Plunien}. We
suppose that the boundary performs harmonic oscillations at the modulation
frequency $\eta $. Then the instantaneous cavity eigenfrequency depends on
time as $\omega _{t}=\omega _{0}+\varepsilon \sin (\eta t)$, where $%
\varepsilon $ is the small modulation amplitude.
Normalizing the unperturbed cavity frequency to $\omega _{0}=1$, we write
the modulation frequency as $\eta =2\left( 1+x\right) $, where $x$ is a
small resonance shift. For a weak modulation, $|\varepsilon |\ll 1$, we can
write to the first order in $\varepsilon $: $\chi _{t}\equiv (4\omega
_{t})^{-1}d\omega _{t}/dt\simeq 2q\cos (\eta t)$ \cite{PLA,Law94,Plunien},
where $q\equiv \varepsilon \left( 1+x\right) /4$. Moreover, the term $\omega
_{t}n$ in $H_0$ can be replaced simply by $n$, as soon as the main
effect of modulation is due to the presence of operators $a^{2}$ and $%
a^{\dagger 2}$ in the squeezing part of $H_0$, but not due to the
photon number preserving part $\omega _{t}a^{\dagger }a$.

In the empty cavity, the resonance generation of many photons is achieved
for $x=0$ (being impossible if $|x| \gtrsim |\varepsilon|$ \cite{D98}). On
the other hand, it was shown \cite{PLA} that no more than two photons can be
created in the presence of a single atom if $|\varepsilon| \ll |g_1|$, and
this can happen if $|x| \sim |g_1|$. Our aim is to find the resonance
regimes in the presence of \emph{two\/} atoms for different relations
between the parameters $\varepsilon$, $g_j$ and $\Omega_j$. We show that
there are two types of resonances. For some distinguished values of $x \neq
0 $ at most two photons can be created. But under certain conditions, the
multiphoton generation becomes possible again for $x \approx 0$ (contrary to
the one-atom case), even if $|\varepsilon| \ll |g_1|$. This interesting
result is one of the main motivations for this publication.

The dynamics of the closed system (atoms + field mode) is governed
(neglecting dissipation)
by the Schr\"{o}dinger equation $i\partial |\Psi
(t)\rangle /\partial t=H_{0}|\Psi (t)\rangle $.
To find analytical solutions we go to the \emph{interaction
picture\/}:  
$
|\Psi(t)\rangle = \exp \left[-it\left( \eta /2\right)
\left( n+\sigma _{1}^{z}/2+\sigma _{2}^{z}/2\right) \right] |\psi (t)\rangle$,
 since the Hamiltonian acting upon the new wavefunction $|\psi(t)\rangle $
 becomes time independent after the Rotating Wave Approximation
(RWA):
\[
H_{I} =\sum_{j=1}^{2}\Big[ g_{j}(a\sigma _{j}^{+}+a^{\dagger }\sigma _{j}^{-})
-\frac{\Delta _{j}+x}{2}\sigma_{j}^{z}\Big]
-iq( a^{2}-a^{\dagger 2}) -xn,
\]%
where $ \Delta _{j}=1-\Omega _{j}$.
We expand the wavefunction in the atom and Fock bases as follows:
\begin{widetext}
\begin{eqnarray}
|\psi(t) \rangle &=&\sum_{m=0}^{\infty} e^{ixmt}\left[ a_{m}(t)e^{-i\left(
2x+\Delta _{1}+\Delta _{2}\right) t/2}|\overline{g}_{1}\rangle|\overline{g}%
_{2}\rangle|m\rangle +b_{m}(t)e^{-i\left( \Delta _{1}-\Delta _{2}\right)
t/2}|\overline{g}_{1}\rangle|\overline{e}_{2}\rangle|m\rangle \right.  \notag
\\
&&\left. +c_{m}(t)e^{i\left( \Delta _{1}-\Delta _{2}\right) t/2}|\overline{e}%
_{1}\rangle|\overline{g}_{2}\rangle|m\rangle +d_{m}(t)e^{i\left( 2x+\Delta
_{1}+\Delta _{2}\right) t/2}|\overline{e}_{1}\rangle|\overline{e}%
_{2}\rangle|m\rangle \right] .
\end{eqnarray}%
%
Then the Schr\"{o}dinger equation with Hamiltonian $H_{I}$ leads to the
set of coupled differential equations%
\begin{eqnarray}
\dot{a}_{m} &=&-ig_{1}\sqrt{m}c_{m-1}e^{i\Delta _{1}t}-ig_{2}\sqrt{m}%
b_{m-1}e^{i\Delta _{2}t}+q\hat{W}_{m}a_{m}  \label{a1} \\
\dot{b}_{m-1} &=&-ig_{1}\sqrt{m-1}d_{m-2}e^{i\Delta _{1}t}-ig_{2}\sqrt{m}%
a_{m}e^{-i\Delta _{2}t}+q\hat{W}_{m-1}b_{m-1} \\
\dot{c}_{m-1} &=&-ig_{1}\sqrt{m}a_{m}e^{-i\Delta _{1}t}-ig_{2}\sqrt{m-1}%
d_{m-2}e^{i\Delta _{2}t}+q\hat{W}_{m-1}c_{m-1} \\
\dot{d}_{m-2} &=&-ig_{1}\sqrt{m-1}b_{m-1}e^{-i\Delta _{1}t}-ig_{2}\sqrt{m-1}%
c_{m-1}e^{-i\Delta _{2}t}+q\hat{W}_{m-2}d_{m-2},  \label{a4}
\end{eqnarray}%
where $\hat{W}_{m}O_{m}\equiv \sqrt{m\left( m-1\right) }O_{m-2}e^{-2ixt}-%
\sqrt{\left( m+1\right) \left( m+2\right) }O_{m+2}e^{2ixt}$. %
\end{widetext}

\emph{Weak modulation with atoms in resonance}.--This regime
is defined by the inequality $|\varepsilon |\ll G\equiv \sqrt{%
g_{1}^{2}+g_{2}^{2}}$.
If two atoms are in resonance, $\Delta _{1}=\Delta _{2}=0$, the solution
to Eqs. (\ref{a1})-(\ref{a4}) in the absence of external modulation ($q=0$)
is (for $m\geq 2$)%
\begin{eqnarray}
a_{m} &=&\sum_{\alpha ,\beta =+,-}\mathcal{F}_{m}^{\alpha \beta }\exp
(\alpha iGL_{m}^{\beta }t),  \label{w1} \\
d_{m-2} &=&-\sum_{\alpha ,\beta =+,-}V_{m}^{\beta }\mathcal{F}_{m}^{\alpha
\beta }\exp (\alpha iGL_{m}^{\beta }t), \\
b_{m-1} &=&\frac{G}{\left( g_{1}^{2}-g_{2}^{2}\right) }\sum_{\alpha ,\beta
=+,-}\alpha L_{m}^{\beta }\mathcal{F}_{m}^{\alpha \beta }\exp (\alpha
iGL_{m}^{\beta }t)  \notag \\
&&\times \left[ g_{2}/\sqrt{m}+g_{1}V_{m}^{\beta }/\sqrt{m-1}\right] , \\
c_{m-1} &=&b_{m-1}[g_{1}\rightarrow g_{2};g_{2}\rightarrow g_{1}],
\label{w4}
\end{eqnarray}%
where $\mathcal{F}_{m}^{\alpha \beta }$ are constant coefficients,
\begin{eqnarray*}
V_{m}^{\pm } &=&\frac{1\mp 2R_{m}}{2\rho \sqrt{m\left( m-1\right) }}\,,\quad
\rho =\frac{2g_{1}g_{2}}{G^{2}}, \\
R_{m} &=&\frac{1}{2}\sqrt{1+4\rho ^{2}m\left( m-1\right) },\quad L_{m}^{\pm
}=\sqrt{m-{1}/{2}\pm R_{m}}\,.
\end{eqnarray*}%
Substituting now expressions (\ref{w1})-(\ref{w4}) back into Eqs. (\ref{a1}%
)-(\ref{a4}) and assuming that $\mathcal{F}_{m}^{\alpha \beta }$ are slowly
varying functions of time, one can verify that for specific values of the
resonance shift $x$ some of these functions become multiplied by imaginary
exponentials with large arguments (compared to $q$), while others are
multiplied by time-independent coefficients, so one is allowed to perform
the RWA and obtain simplified effective dynamics. We find that for the
initial zero-excitation state $|\overline{g}_{1}\rangle |\overline{g}%
_{2}\rangle |0\rangle $ at most two photons can be created whenever $%
G\left\vert L_{4}^{\pm }-L_{2}^{\pm }\right\vert \gg q$. The resonant
regimes occur for $2x=-\alpha GL_{2}^{\beta }$ (with $\alpha ,\beta =+,-$),
when the only nonzero amplitudes (neglecting small terms of the order of $%
\varepsilon /G$) are $a_{0}=\cos (qtR_{\beta })$ (it does not depend on the
sign of $\alpha $) and $\mathcal{F}_{2}^{\alpha \beta }=R_{\beta }\sin
(qtR_{\beta })/\sqrt{2}$, where $R_{\pm }=\frac{1}{2}\sqrt{2\pm R_{2}^{-1}}$.

For a single atom ($g_2=0$) one has $R_m \equiv 1/2$, so that $R_{+}= 1$ and
$R_{-}=0$. Then the only resonances with a periodic creation of at most two
photons happen for $x= \pm |g_1|/\sqrt{2}$. In this case $a_{0}=\cos (qt)$,
while the only other nonzero coefficients are $\mathcal{F}_2^{\mp +}=
\sin(qt)/\sqrt{2}$ in accordance with \cite{PLA}. In the presence of the
second atom, new resonances become possible. If $|g_2| \ll |g_1|$, then
these additional resonance frequencies have the values $x \approx \pm
|g_1|/2 $. However, since $R_{-}\approx \rho\sqrt{2} \ll 1$ in this case,
the corresponding dynamics is quite slow and the probability of the photon
creation is small, too.

\begin{figure}[htb]
\begin{center}
\includegraphics[width=0.49\textwidth]{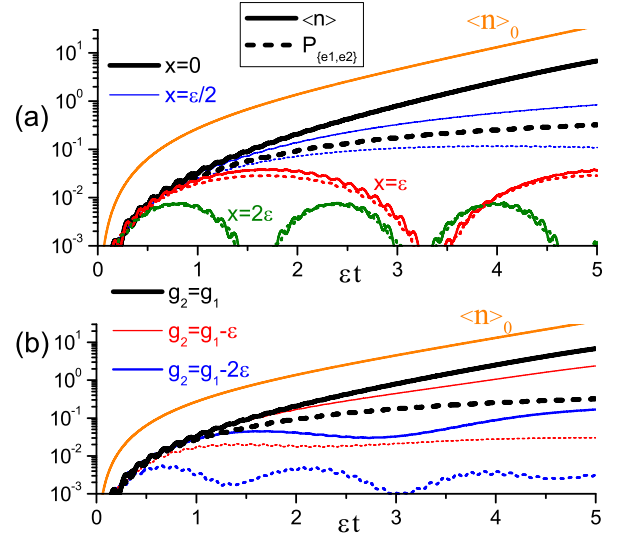} {}
\end{center}
\par
\vspace{-1cm}
\caption{(Color online) The mean photon number (solid lines) and atomic
excitation probabilities (dashed lines) as functions of dimensionless time $%
\protect\varepsilon t$. \textbf{(a)} The influence of nonzero detuning $x$
for $g_1=g_2$. \textbf{(b)} The influence of disbalance $g_2-g_1$ for $x=0$.
Numerical values of parameters are given in the text.}
\label{fig-x0}
\end{figure}

The most interesting situation takes place if $|g_1|=|g_2|$. Then $R_m
=m-1/2 $ and $L_m^{-} \equiv 0$. We still have the resonances at $x= \pm
|g_1| \sqrt{3/2}$, when no more than two photons can be created from the
initial ground state, since the only nonzero coefficients in this case are $%
a_{0}=\cos (\sqrt{2/3}qt)$ and $\mathcal{F}_2^{\mp +} = r\sin (\sqrt{2/3}qt)/%
\sqrt{3}$, where $r=g_2/g_1 = \pm 1$. But two other resonances merge in the
single one at $x=0$. In this case, solving Eqs. (\ref{a1})-(\ref{a4}) with $%
q=0$, one can write (for $m\geq 2$)%
\begin{eqnarray*}
a_{m} &=& r\left[\mathcal{W}_{m}E_m^{-}(t)+\mathcal{X}_{m}E_m^{+}(t)+%
\mathcal{Y}_{m}\right], \\
b_{m-1} &=&\sqrt{1-(2m)^{-1}}\left[\mathcal{W}_{m}E_m^{-}(t) -\mathcal{X}%
_{m}E_m^{+}(t)\right]+\mathcal{Z}_{m}, \\
c_{m-1} &=&r\left(b_{m-1}-2\mathcal{Z}_{m}\right), \\
d_{m-2} &=& r a_m\sqrt{\frac{m-1}{m}} -\frac{2m-1}{\sqrt{m(m-1)}}\mathcal{Y}%
_{m},
\end{eqnarray*}
where $E_m^{\pm}(t)=\exp[\pm ig_1\sqrt{2(2m-1)}\,t]$.
In the presence of additional terms proportional to the small parameter $%
q\ll G$ in Eqs. (\ref{a1})-(\ref{a4}), the coefficients $\mathcal{W}_{m}$, $%
\mathcal{X}_{m}$, $\mathcal{Y}_{m}$ and $\mathcal{Z}_{m}$ become
time-dependent. For the standard atomless DCE resonance $\eta =2$, assuming
that $\left\vert \mathcal{W}_{m}\right\vert ,\left\vert \mathcal{X}%
_{m}\right\vert \ll 1$ for all $m$, we perform the RWA and find that $%
\mathcal{Z}_{m}(t)=0$, meaning that $b_{m}(t),c_{m}(t)\approx 0$
for all times. Only functions $\mathcal{Y}_{m}$ vary slowly with time according
to the equations%
\begin{eqnarray}
\mathcal{\dot{Y}}_{m} &\simeq &q\left[ \sqrt{m(m-1)}\,\frac{2m-3}{2m-1}%
\mathcal{Y}_{m-2}\right.   \notag \\
&-&\left. \sqrt{(m+1)(m+2)}\,\frac{m-1}{m+1}\frac{2m+1}{2m-1}\mathcal{Y}_{m+2}%
\right]   \label{bt}
\end{eqnarray}%
with the initial condition $\mathcal{Y}_{m}(0)= r\delta_{m0}$.
Therefore eventually all (even) coefficients $\mathcal{Y}_{m}$ become
different from zero, so that many photons can be created from the
initial vacuum state. Eq. (\ref{bt})
has two remarkable properties. First, it does not contain
the atomic coupling coefficients.
Second, the fractions in its right-hand side tend to the unit values for
$m\gg 1$, and in this limit Eq. (\ref{bt}) has the same form as the
equation governing the evolution of the field amplitudes (in the Fock basis)
in the empty cavity. Since the main contribution to the mean photon number
$\langle n(t)\rangle$ is given by the
coefficients $\mathcal{Y}_{m}$ with $m\gg 1$ if $\langle n\rangle \gg 1$,
we can expect that after some transient time
the photons will be steadily created with the same asymptotical rate
$d\ln (\langle n\rangle )/d(\varepsilon t)$ as in the empty cavity.
Moreover, since $\left\vert d_{m-2}(t)\right\vert
^{2}=[m/(m-1)]\,\left\vert a_{m}(t)\right\vert ^{2}$, both atoms become
excited simultaneously. Numerical calculations confirm these predictions, as
shown in Fig. \ref{fig-x0}, where we plot the mean photon number $%
\left\langle n\right\rangle $ and the probability of double excitation $%
P_{\{e1,e2\}}$ for parameters $g_{1}=4\times 10^{-2}$ and $\varepsilon
=2\times 10^{-3}$
\footnote{All numerical calculations have been performed for the initial
Hamiltonian $H_0$ without any simplifications. The scheme of such calculations
was described briefly in \cite{Resdyn}.
We verified that the analytical results according to Eq. (\ref{bt}) are
indistinguishable from the numerical ones within the thicknesses of lines.
}.
Part (a) shows the role of the detuning parameter $x$
when $g_{2}=g_{1}$: the photon creation and atomic excitations
practically stop for $x\gtrsim \varepsilon $.
Part (b) shows the influence of disbalance $g_{2}-g_{1}$ when $x=0$:
again, all effects practically disappear if $|g_{2}-g_{1}|%
\gtrsim \varepsilon $.

The mean number of photons for $x=0$ is smaller than
that in the empty-cavity case,
$\langle n_0(t)\rangle = \sinh^2\left(\varepsilon t/2\right)$,
due to initial transient processes, when
the atomic populations attain stationary values:
one can see that
 the line $\langle n(t)\rangle$ can be obtained from
$\langle n_0(t)\rangle$ by some positive shift in time.
Therefore the $x=0$ resonance for $\left\vert
g_{1}\right\vert =\left\vert g_{2}\right\vert $ is interesting from the
point of view of detecting Casimir photons, since the atoms get excited
simultaneously without interrupting the photon generation process.

If the second atom is in the dispersive regime, $|g_{2}|\ll \left\vert
\Delta _{2}\right\vert $ (while $\Delta_1=0$), we define the dispersive
shift $\delta _{2}\equiv g_{2}^{2}/\Delta _{2}$ and repeating the previous
steps we find that for $\left\vert \delta _{2}\right\vert \ll |g_{1}|$ the
photon generation occurs for the resonance shifts $2x=\left( 3/2\right)
\delta _{2} \pm G_{2}$ with $G_{2}\equiv \sqrt{2g_{1}^{2}+\delta _{2}^{2}/4}$. The resulting nonzero probability amplitudes read: $a_{0} =\cos \left( qt\sqrt{1\pm\delta _{2}/(2G_{2})}\right)$,%
\begin{eqnarray*}
a_{2} &=&e^{-i\left( 3/2\right) \delta _{2}t}\left[ \mathcal{W}e^{-iG_{2}t}+%
\mathcal{X}e^{iG_{2}t}\right] , \\
c_{1} &=&\frac{G_{2}e^{-i\left( 3/2\right) \delta _{2}t}}{\sqrt{2}g_{1}}%
\left\{ \mathcal{W}\left[ 1-\delta _{2}/(2G_{2})\right] e^{-iG_{2}t}\right.
\\
&&\left. -\mathcal{X}\left[ 1+\delta _{2}/(2G_{2})\right] e^{iG_{2}t}\right%
\} , \\
b_{1} &\simeq &\sqrt{2}(g_{2}/\Delta _{2})e^{-i\Delta _{2}t}a_{2},\quad
d_{0}\simeq (g_{2}/\Delta _{2})e^{-i\Delta _{2}t}c_{1}, \\
\binom{\mathcal{W}}{\mathcal{X}}&=&\frac{\sqrt{1\pm\delta _{2}/(2G_{2})}}{%
\sqrt{2}}\sin \left( qt\sqrt{1\pm\delta _{2}/(2G_{2})}\right).
\end{eqnarray*}%
At most two photons can be created in this case.

\emph{Dispersive regimes}.--Many photons can be generated from vacuum if
both atoms are in the dispersive regime, $|g_{j}|\ll \left\vert \Delta
_{j}\right\vert $. In this case, instead of solving coupled differential
equations it is convenient to write the wavefunction $|\psi (t)\rangle $ as
\cite{Resact,Tridin}
$
|\psi (t)\rangle =U^{\dagger }\exp \left( -iH_{ef}t\right) U|\psi (0)\rangle$
,
where the effective Hamiltonian $H_{ef}\equiv UH_{I}U^{\dagger }$ is defined
by means of the unitary operator $U=\exp (Y)$. Choosing $Y=a^{\dagger
}\left( \zeta _{2}\sigma _{2}^{-}+\zeta _{1}\sigma _{1}^{-}\right) -h.c.$
(where $\zeta _{j}=g_{j}/\Delta _{j}$ are small parameters, $\left\vert
\zeta _{j}\right\vert \ll 1$, $j=1,2$)
and expanding the exponentials in Taylor's series we obtain to the second
order in $\zeta _{j}$ [assuming $\mathcal{O}(\zeta _{1})\sim \mathcal{O}%
(\zeta _{2})$]%
\begin{eqnarray}
H_{ef} &=&-(x+\delta _{1}\sigma _{1}^{z}+\delta _{2}\sigma
_{2}^{z})n-\sum_{j=1}^{2}\frac{\Delta _{j}+x+\delta _{j}}{2}\sigma _{j}^{z}
\notag \\
&&-\zeta _{1}\zeta _{2}\left[ \frac{\Delta _{1}+\Delta _{2}}{2}\sigma
_{1}^{+}\sigma _{2}^{-}-2iq\sigma _{1}^{+}\sigma _{2}^{+}+h.c.\right]  \notag
\\
&&-iq\left[ \left( 1+\zeta _{1}^{2}\sigma _{1}^{z}+\zeta _{2}^{2}\sigma
_{2}^{z}\right) a^{2}-h.c.\right] .  \label{eH}
\end{eqnarray}%
Here $\delta _{j}=g_{j}^{2}/\Delta _{j}$ are the dispersive shifts ($j=1,2$%
). In view of the perturbative expansion the effective Hamiltonian (\ref{eH}%
) is valid roughly for times $\left\vert \delta _{1}\right\vert t\ll 1$. For
$q=0$ it describes the indirect interaction between the two atoms via the
cavity field \cite{Majer}. Since the state $|\overline{g}_{1}\rangle |%
\overline{g}_{2}\rangle |0\rangle $ is the common eigenstate of $Y$ and $%
(\sigma _{1}^{+}\sigma _{2}^{-}+h.c.)$ with null eigenvalues, one has $U|%
\overline{g}_{1}\rangle |\overline{g}_{2}\rangle |0\rangle =|\overline{g}%
_{1}\rangle |\overline{g}_{2}\rangle |0\rangle $, so the term $(\sigma
_{1}^{+}\sigma _{2}^{-}+h.c.)$ can be dropped out
if $|\psi (0)\rangle =|\overline{g}_{1}\rangle |\overline{g}_{2}\rangle
|0\rangle $. Besides, if the coefficient in front of $n$ in Eq. (\ref{eH})
is adjusted to zero, the photon generation term $iq\left( 1+\zeta
_{1}^{2}\sigma _{1}^{z}+\zeta _{2}^{2}\sigma _{2}^{z}\right) a^{\dagger 2}$
becomes resonant, while the term $2iq\zeta _{1}\zeta _{2}(\sigma
_{1}^{+}\sigma _{2}^{+}-h.c.)$ can be neglected for initial times. In this
case the wavefunction for the resonance shift $x=\delta _{1}+\delta _{2}$
reads as (neglecting a global phase)
$
|\psi (t)\rangle =U^{\dagger }\hat{\Lambda}\left( 1-\zeta ^{2}\right) |%
\overline{g}_{1}\rangle |\overline{g}_{2}\rangle |0\rangle $,
where
the squeezing operator $\hat{\Lambda}(v)\equiv \exp [v\,qt(a^{\dagger
2}-a^{2})]$ has the property \cite{Puri,Resact}
$\hat{\Lambda}^{\dagger }(v)a\hat{\Lambda}(v)=%
\mathcal{C}_{v}a+\mathcal{S}_{v}a^{\dagger }$, with $\mathcal{C}_{v}=\cosh
\left( 2vqt\right) $, $\mathcal{S}_{v}=\sinh \left( 2vqt\right) $,
and $ \zeta^{2}\equiv \zeta _{1}^{2}+\zeta _{2}^{2}$.

\begin{figure}[htb]
\begin{center}
\includegraphics[width=0.49\textwidth]{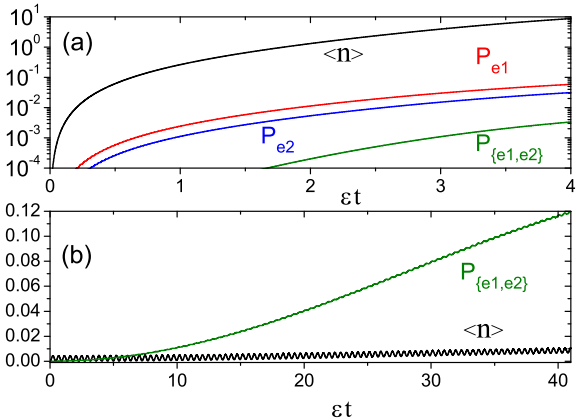} {}
\end{center}
\par
\vspace{-1cm}
\caption{(Color online) The mean photon number and atomic excitation
probabilities versus the dimensionless time $\protect\varepsilon t$ in
the dispersive regimes: \textbf{(a)} $x=\protect\delta _{1}+\protect\delta %
_{2}$; \textbf{(b)} $2x=-\sum_{j=1}^{2}\left( \Delta _{j}+\protect\delta %
_{j}\right) $. Other parameters are specified in the text.}
\label{f2}
\end{figure}

Average values of the main observable quantities are as follows (to the
second order in $\zeta _{j}$):
\begin{eqnarray*}
\left\langle n(t)\right\rangle &=&\left( 1-\zeta^{2}\right) \sinh ^{2}\left[
2qt\left( 1-\zeta^{2}\right) \right] , \\
P_{e1}(t) &=&\zeta _{1}^{2}\left\langle n(t)\right\rangle , \quad
P_{e2}(t)=\zeta _{2}^{2}\left\langle n(t)\right\rangle , \\
\langle \left( \Delta X_{\pm }\right) ^{2}\rangle &=&\frac12\left\{\zeta^{2}
+\left(1-\zeta^{2}\right)\exp\left[\pm 4qt\left(1-\zeta^{2}\right)\right]
\right\},
\end{eqnarray*}
where $X_{+}=\left( a+a^{\dagger }\right) /\sqrt{2}$ and $X_{-}=\left(
a-a^{\dagger }\right) /(\sqrt{2}i)$ are the field quadratures. Moreover, for
times $\left\vert \delta _{1}\right\vert t\ll 1$ the probability $%
P_{\{e1,e2\}}$ of detecting simultaneously both atoms in their excited
states is proportional to $\zeta _{1}^{4}$, so it is very small. Therefore,
by measuring $P_{e1}$ or $P_{e2}$ one can estimate the mean photon number.
In Fig. \ref{f2}a we show the behavior of $\left\langle n\right\rangle $, $%
P_{e1}$, $P_{e2}$ and $P_{\{e1,e2\}}$ for parameters
$\varepsilon =2\times 10^{-3}$,
$g_{1}=4\times 10^{-2}$%
, $g_{2}=3\times 10^{-2}$, $\Delta _{1}=10g_{1}$, $\Delta _{2}=15g_{2}$,
and $x=\delta _{1}+\delta _{2}$. We see
that many photons are created and the atomic populations are proportional to
the mean photon number, while the probability of double atomic excitation is
very small.

If $\left\vert \sum_{j=1}^{2}\left( \Delta _{j}+3\delta _{j}\right)
\right\vert \gg q$ and the resonance shift
is tuned to $2x=-\sum_{j=1}^{2}\left( \Delta _{j}+\delta _{j}\right) $
with $\Delta _{1}\sim -\Delta _{2}$, then
the photon generation term becomes off-resonant and the only resonant term $%
2iq\zeta _{1}\zeta _{2}(\sigma _{1}^{+}\sigma _{2}^{+}-h.c.)$ survives in
the interaction part of the effective Hamiltonian (\ref{eH}) even to higher
orders in $\zeta _{1}$, extending its validity beyond the previous condition
$\left\vert \delta _{1}\right\vert t\ll 1$. In this case only the atomic
excitations are generated at a rather small rate $2q\zeta _{1}\zeta _{2}$
and the probability of detecting both atoms simultaneously in the excited
states is $\left( 1-\zeta _{1}^{2}-\zeta _{2}^{2}\right) \sin ^{2}(2qt\zeta
_{1}\zeta _{2})$. In Fig. \ref{f2}b we show the behavior of $\left\langle
n\right\rangle $ and $P_{\{e1,e2\}}$ for parameters $g_{1}=4\times 10^{-2}$,
$g_{2}=3\times 10^{-2}$, $\Delta _{1}=0.22$, $\Delta _{2}=-0.2$, $%
\varepsilon =2\times 10^{-3}$ and $2x=-\sum_{j=1}^{2}\left( \Delta
_{j}+\delta _{j}\right) $, where we see that double atomic excitations are
created while essentially the field remains in the vacuum state.

\emph{Other regimes}.--If atom 1 is resonant ($\Delta _{1}=0$) and
weakly coupled to the field ($|g_{1}|\ll \varepsilon $), while atom 2 is
in the dispersive regime ($|g_{2}|\ll \left\vert \Delta _{2}\right\vert $),
then we make the transformation with $Y=a^{\dagger }\left( \zeta _{2}\sigma
_{2}^{-}+i\xi _{1}\sigma _{1}^{+}\right) -h.c.$, $\xi _{1}=g_{1}/(2q)$
and $\zeta _{2}=g_{2}/\Delta _{2}$. For the resonance shift $x=\delta _{2}$
the effective Hamiltonian describing parametric amplification reads (after
RWA)%
\begin{eqnarray*}
H_{ef} &=&
-\frac{\Delta _{2}+2\delta _{2}}{2}\sigma _{2}^{z}-iq\left[ \left( 1+\xi
_{1}^{2}\sigma _{1}^{z}+\zeta _{2}^{2}\sigma _{2}^{z}\right) a^{2}-h.c.%
\right]
 \\
&& -\delta _{2}\left( 1-2\xi _{1}^{2}\sigma _{1}^{z}+\sigma
_{2}^{z}\right) n-\frac{\delta _{2}}{2}\left( 1-2\xi _{1}^{2}\right) \sigma
_{1}^{z}.
\end{eqnarray*}%
For the initial state $|\overline{e}_{1}\rangle |\overline{g}_{2}\rangle
|0\rangle $ one has $U|\overline{e}_{1}\rangle |\overline{g}_{2}\rangle
|0\rangle =|\overline{e}_{1}\rangle |\overline{g}_{2}\rangle |0\rangle $, so
 $|\psi (t)\rangle =U^{\dagger }\hat{%
\Lambda}(1+\xi _{1}^{2}-\zeta _{2}^{2})|\overline{e}_{1}\rangle |\overline{g}%
_{2}\rangle |0\rangle $ (up to a global phase). This yields the following average values:%
\begin{eqnarray*}
\left\langle n(t)\right\rangle &=&\left( 1-\xi _{1}^{2}-\zeta
_{2}^{2}\right) \sinh ^{2}\left[ 2q(1+\xi _{1}^{2}-\zeta _{2}^{2})t\right] ,
\\
P_{g1} &=&\xi _{1}^{2}\left\langle n(t)\right\rangle ,\quad P_{e2}=\zeta
_{2}^{2}\left\langle n(t)\right\rangle , \\
\langle \left( \Delta X_{\pm }\right) ^{2}\rangle &=&\frac{\xi
_{1}^{2}+\zeta _{2}^{2}}{2}+\frac{1-\xi _{1}^{2}-\zeta _{2}^{2}}{2}e^{\pm
4q(1+\xi _{1}^{2}-\zeta _{2}^{2})t},
\end{eqnarray*}%
where $P_{g1}$ is the ground state probability of atom 1. Besides, the
probability $P_{\{g1,e2\}}$ of finding simultaneously atom 1 in the
ground state and atom 2 in the excited state is zero [to the second order in
$\mathcal{O}(\xi _{1})$, $\mathcal{O}(\zeta _{2})$].

Analogously, if both atoms are weakly coupled to the field, $G \ll
|\varepsilon| $, then by performing the transformation with $Y=ia^{\dagger
}\left( \xi _{1}\sigma _{1}^{+}+\xi _{2}\sigma _{2}^{+}\right) -h.c.$ and $%
\xi _{j}=g_{j}/(2q)$ one obtains for $x=\Delta _{1}=\Delta _{2}=0$ the
effective Hamiltonian [to the second order in $\xi _{j}$, for $\mathcal{O}%
(\xi _{1})\sim \mathcal{O}(\xi _{2})$]%
\begin{equation*}
H_{ef}=iq\left[ \left( 1+\xi _{1}^{2}\sigma _{1}^{z}+\xi _{2}^{2}\sigma
_{2}^{z}\right) a^{\dagger 2}-2\xi _{1}\xi _{2}\sigma _{1}^{+}\sigma
_{2}^{+}-h.c.\right].
\end{equation*}%
In these cases many photons can be created as well, and the atoms may serve
to monitor the photon generation.

\emph{Conclusions}.--We found that the two-atom nonstationary cavity QED is
attractive from the point of view of producing different types of entangled
states and detecting the DCE, because in specific regimes the atoms can
acquire independent information about the field state without inhibiting the
photon generation process. In particular, we showed that in the realistic
case when the external modulation amplitude is much smaller than the
atom-cavity coupling strengths, many photons, as well as atomic excitations,
can be generated from the initial zero-excitation state even if both atoms
are resonant with the unperturbed cavity field, contrary to the single
2-level atom scenario. Moreover, simply by adjusting the modulation
frequency, keeping the other parameters unaltered, one can achieve the
regime in which at most two photons are generated. If the atoms are
off-resonant, then for the zero-excitation initial state many photons can be
created for a specific modulation frequency; yet by appropriately tuning the
modulation frequency one can achieve the regime in which only atomic
excitations are generated. Furthermore, one can explore the regime in which
one atom is resonant but weakly coupled to the field, while the other atom
is in the dispersive regime -- in this case many photons can be created from
vacuum and the atoms monitor independently the process.
This variety of possibilities can be useful for choosing optimal schemes
of detecting the Casimir photons.
In view of the results obtained, generalizations to the systems of three and
more atoms could be quite interesting. But we leave this problem for another study.

\begin{acknowledgments}
A.V.D. acknowledges the partial support of DPP/UnB. V.V.D. acknowledges the partial support of CNPq (Brazilian agency).
\end{acknowledgments}

\end{document}